\documentclass[prodmode]{acmsmall} 

\usepackage{etex}
\reserveinserts{28}

\usepackage[caption=false]{subfig}
\usepackage{amssymb,amstext,amsmath,amsfonts}
\usepackage[ruled,norelsize]{algorithm2e}
\usepackage{algorithmic}

\usepackage{epsfig,graphicx}
\usepackage{url}
\usepackage{multicol,multirow}
\usepackage{tikz}
\usepackage{pgfplots}
\usepackage[utf8]{inputenc} 
\usepackage{balance}
\usepackage{nicefrac}
\usepackage{xspace}
\usepackage{acronym}
\usepackage{bbm}
\usepackage{xstring}
\usepackage{flushend}

\usepackage{tabularx}
\usepackage{ltxtable}
\usepackage{booktabs}

\usepackage{color}

\newcommand{\promed}{ProMED-mail}
\newcommand{\subparagraph}{}

\newcommand{\meco}{\mbox{\normalfont \textsc{M-Eco}}}

\pgfplotscreateplotcyclelist{bw}{%
dashed, every mark/.append style={solid, fill=gray},mark=diamond*\\%
densely dotted, every mark/.append style={solid, fill=gray}, mark=square*\\%
solid, every mark/.append style={solid, fill=gray}, mark=otimes*\\%
dotted, every mark/.append style={solid, fill=gray}, mark=*\\%
loosely dotted, every mark/.append style={solid, fill=gray}, mark=triangle*\\%
loosely dashed, every mark/.append style={solid, fill=gray},mark=*\\%
densely dashed, every mark/.append style={solid, fill=gray},mark=square*\\%
dashdotted, every mark/.append style={solid, fill=gray},mark=otimes*\\%
dasdotdotted, every mark/.append style={solid},mark=star\\%
densely dashdotted,every mark/.append style={solid, fill=gray},mark=diamond*\\%
}

\usepackage[ruled]{algorithm2e}
\renewcommand{\algorithmcfname}{ALGORITHM}
\SetAlFnt{\small}
\SetAlCapFnt{\small}
\SetAlCapNameFnt{\small}
\SetAlCapHSkip{0pt}
\IncMargin{-\parindent}

\acmYear{2016}

\begin{document}

\title{Why is it Difficult to Detect Sudden and Unexpected\\Epidemic Outbreaks in Twitter?}

\markboth{Avar\'{e} Stewart et al.}{Why is it Difficult to Detect Sudden and Unexpected Epidemic Outbreaks in Twitter?}

\author{AVAR\'{E} STEWART
\affil{L3S Research Center, Leibniz University of Hannover, Germany}
SARA ROMANO
\affil{University of Naples Federico II, Italy}
NATTIYA KANHABUA
\affil{L3S Research Center, Leibniz University of Hannover, Germany}
SERGIO DI MARTINO
\affil{University of Naples Federico II, Italy}
WOLF SIBERSKI
\affil{L3S Research Center, Leibniz University of Hannover, Germany}
ANTONINO MAZZEO
\affil{University of Naples Federico II, Italy}
WOLFGANG NEJDL
\affil{L3S Research Center, Leibniz University of Hannover, Germany}
ERNESTO DIAZ-AVILES
\affil{L3S Research Center, Leibniz University of Hannover, Germany}
}

\begin{abstract}
Social media services such as Twitter are a valuable source of information for decision support systems. Many studies have shown that this also holds for the medical domain, where Twitter is considered a viable tool for public health officials to sift through relevant information for the early detection, management, and control of epidemic outbreaks. This is possible due to the inherent capability of social media services to transmit information faster than traditional channels. However, the majority of current studies have limited their scope to the detection of common and seasonal health recurring events (e.g., Influenza-like Illness), partially due to the noisy nature of Twitter data, which makes outbreak detection and management very challenging.
Within the European project \meco, we developed a Twitter-based Epidemic Intelligence (EI) system, which is designed to also handle a more general class of unexpected and aperiodic outbreaks. In particular, we faced three main research challenges in this endeavor:
1)~\emph{dynamic classification} to manage terminology evolution of Twitter messages, 2)~\emph{alert generation} to produce reliable outbreak alerts analyzing the (noisy) tweet time series, and 3)~\emph{ranking and recommendation} to support domain experts for better assessment of the generated alerts. 
In this paper, we empirically evaluate our proposed approach to these challenges using real-world outbreak datasets and a large collection of tweets. We validate our solution with domain experts, describe our experiences, and give a more realistic view on the benefits and issues of analyzing social media for public health.  
\end{abstract}


%


\acmformat{Avar\'{e} Stewart, Sara Romano, Nattiya Kanhabua, Sergio Di Martino, Wolf Siberski, Antonino Mazzeo,  Wolfgang Nejdl, and Ernesto Diaz-Aviles. 2016. Why is it Difficult to Detect Sudden and Unexpected Epidemic Outbreaks in Twitter?}

\begin{bottomstuff}
Author's addresses: A. Stewart, N. Kanhabua, W. Siberski, W. Nejdl, and E.~Diaz-Aviles.\break L3S Research Center, Leibniz University of Hannover, Germany, email: \texttt{\{stewart, kanhabua, siberski, nejdl, diaz\}@L3S.de}; S. Romano, S. Di Martino, {and} A. Mazzeo, DIETI, University of Naples Federico II, Italy, email:\texttt{\{sara.romano, sergio.dimartino, mazzeo\}@unina.it}.
%
\setcopyright{rightsretained}
\doi{}
\end{bottomstuff}

\maketitle

\section{Introduction}
\label{sec:intro}
Public health officials face new challenges for outbreak alert and response due to the continuous emergence of infectious diseases and their contributing factors --~e.g., demographic change and globalization. Only the early detection of disease activity, followed by a rapid response, can reduce the impact of epidemics. Conflictingly, the (slow) speed with which information propagates through the traditional channels can undermine time-sensitive strategies.

Online Social Networks are valuable sources for real-time information such as status updates, opinions, or news in many domains. In particular, Twitter, a free social network that enables its users to post/read short messages called \textit{tweets}, has been shown to be capable of transmitting information faster than traditional media channels for detecting natural disasters~\cite{SakakiOkazaki:2010}, emergency situations~\cite{cassa2013twitter}, political persuasion~\cite{Borge-Holthoefer:2015:CND:2675133.2675163}, or current trends~\cite{LamposTIST2011}.
%
 
In recent years it has been widely recognized that Twitter can also be used as a data source for digital health surveillance. The monitoring of the social stream is intended as an extension or complement to traditional passive surveillance systems~\cite{milinovich2014internet,Neill12}, whose goal is to give public health officials a head start in detecting and managing outbreaks~\cite{DBLP:journals/expert/Dredze12,Kostkova:2014:UTE:2659230.2597892,Oyeyemig6178,Odlum2015563}. As a consequence, in the research community there has been a surge in dealing with tweets related to public health, with a number of proposals for new Epidemic Intelligence (EI) systems. EI has emerged as a type of intelligence gathering aimed to detect the events of interest to public health from unstructured text on the Web. 

Detecting and monitoring outbreak events in Twitter is still challenging due to three main issues: 

	\noindent\textbf{(1) Understanding if a tweet is relevant for outbreak alert.} Putting tweets in the right context for a very broad range of diseases is very difficult, in part, due to the brevity of the tweet messages, which are limited to 140 characters. 
	Although methods for detecting recurring events is mature, the detection of sudden, unexpected, and aperiodic outbreak events requires adaptive approaches to enable the identification of new emergent terms associated to epidemic outbreaks.

	\noindent\textbf{(2) Detecting changes in tweets' time series.} Time series created from tweets are noisy, highly ambiguous and sparse \cite{LamposTIST2011}. Moreover, the characteristics of infectious diseases are highly dynamic in time and space, and their behavior varies greatly. Given this imperfect data, it is important to consider measures for assessing the reliability of alerts, i.e., the extent to which we can actually trust alerts that have been generated for early warning.
	
	\noindent\textbf{(3) Supporting public health officials.} Every day, hundreds of millions of tweets are created world-wide and despite the relatively small fraction of health-related ones, officials still need assistance to cope with the cognitive challenges of exploring a large number of tweets linked to outbreak alerts. The effectiveness of straightforward approaches to retrieval and collaborative filtering can be unsatisfactory, given the dynamics of streaming data and the limited context of detected alerts.

Most studies on the use of Twitter data for outbreak detection have been focused only on the first issue (e.g.,~\cite{paul2012model,WWWDIAZ2012}). In addition, they have been tailored for one or two diseases, thus dealing with a more or less uniform temporal distribution of the tweets. 

Although numerous approaches successfully detect common epidemic outbreak events from Twitter, e.g., seasonal influenza ~\cite{Culotta:2010:TDI:1964858.1964874,DBLP:conf/emnlp/AramakiMM11,LamposTIST2011}, it seems that the challenges in building an EI system are still underestimated, especially when it comes to detecting \textit{emerging} (\textit{novel} or \textit{non-seasonal}) health events from social media streams.

In this work, we propose an event-based EI system that also considers the detection of unexpected and aperiodic public health events. Our goal is to assist officials to retrieve and explore the detected \textit{alerts} for infectious disease outbreaks. This effort represents the outcome of the collaboration with medical domain experts and epidemiologists within the European research project \meco\ -- \textit{Medical Ecosystem: Personalized Event-based Surveillance}~\cite{DBLP:conf/www/DeneckeDS12}. 
In summary, the contributions of this work are as follows: 

\begin{enumerate}
\item We present and empirically evaluate an EI system based on Twitter, which is the result of close collaboration with domain experts and epidemiologists. We experimentally show the effectiveness of our approach to support the task of sudden and unexpected outbreak detection, management and control, and provide insightful lessons for similar endeavors.

\item We propose a novel dynamic classification method for identifying health-related tweets, which is capable to maintain classification accuracy over time.
%

%

\item We conduct a comparative study of surveillance algorithms for alert generation, and present our findings that outline the conditions under which early warnings, generated from Twitter, can be reliable.

\item We present a personalized tweet ranking method for EI, which helps end-users to cope with the cognitive challenges of search and exploration of outbreak alerts.

\end{enumerate}
\section*{M-Eco System Overview}
\label{sec:overview}
%
%
The goal of \meco\ project is to complement and enhance the capabilities of traditional disease surveillance systems. To this end, \meco\ uses novel approaches for early detection and management of emerging threats, and analyzes non-traditional sources such as social media data streams. In addition, \meco\ leverages personalization and filtering techniques to ease the outbreak analysis and control tasks.~\cite{DBLP:conf/www/DeneckeDS12}
%

\begin{figure}[!tb]
	\centering
	\subfloat[]{\label{fig:pipeline}\fbox{\includegraphics[width=0.55\columnwidth]{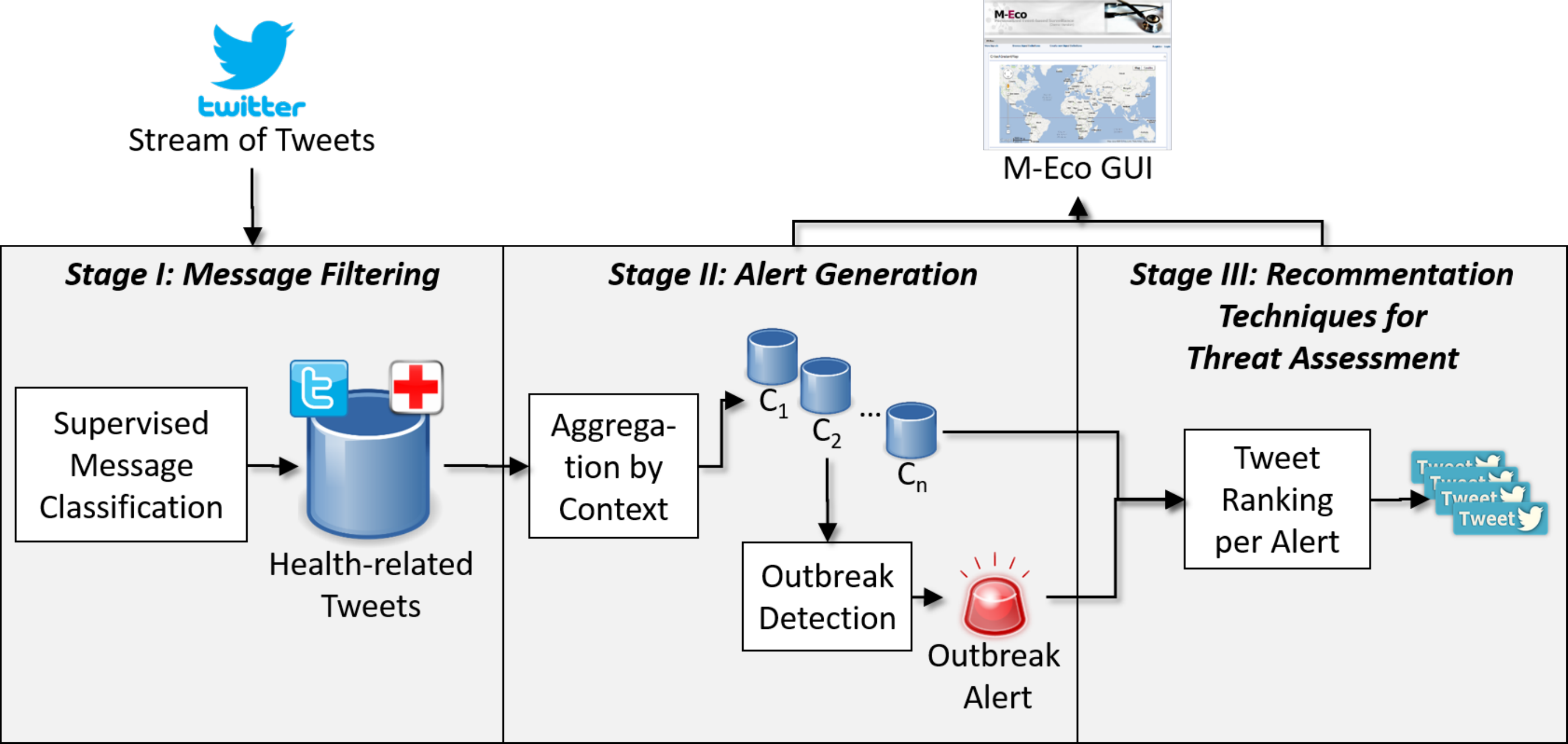}}}
	\quad
	\subfloat[]{\label{fig:GUI}\includegraphics[height=0.215\textheight]{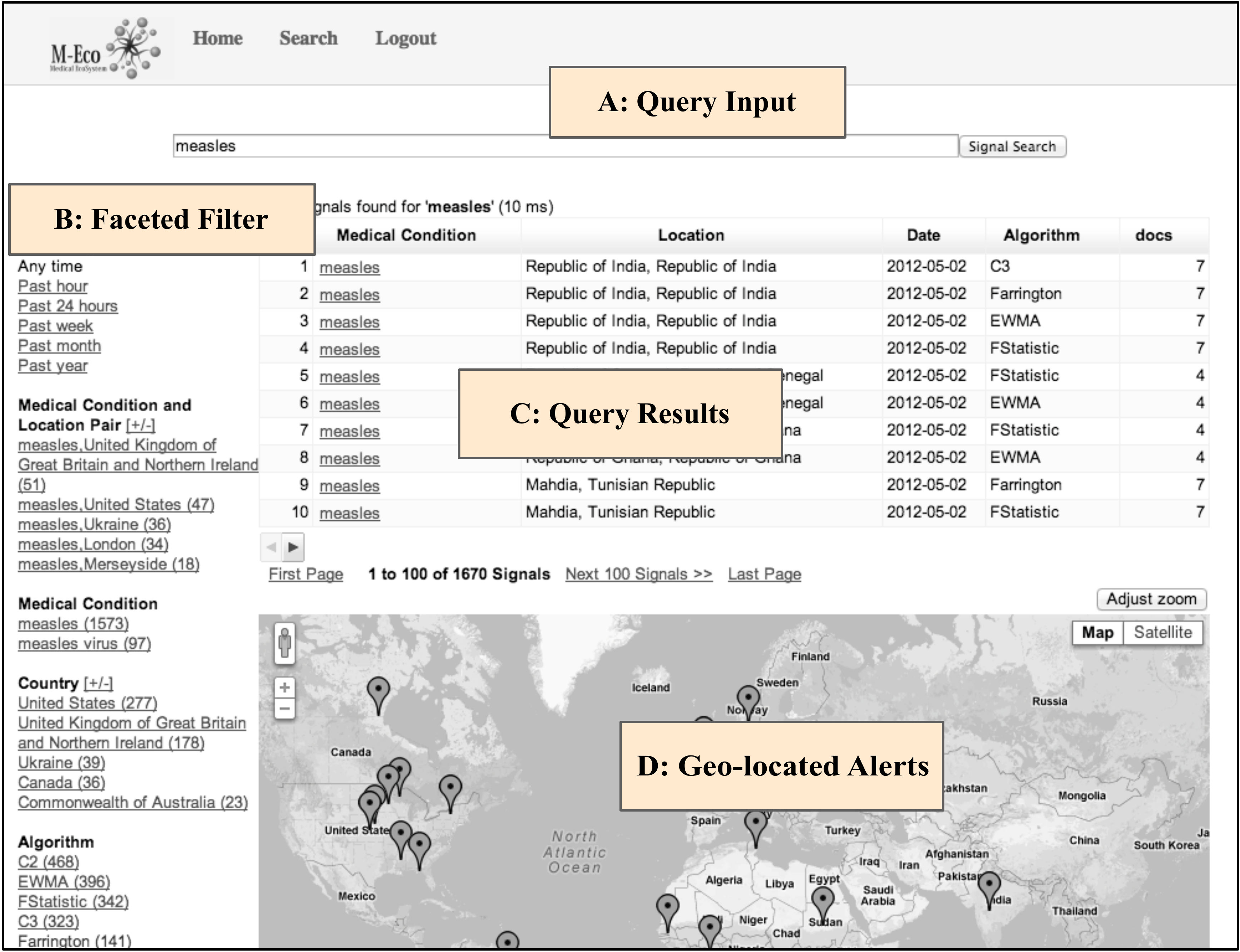}}
	\caption{(a)~Overview of the \meco\ System for Personalized Event-based
		Surveillance; consisting of three main stages of the media stream
		processing pipeline. (b)~\meco\ Alert-based Search Interface consisting of \textit{A. Query Input}: for submitting a search term, e.g., ``measles'', \textit{B. Faceted Filter}: options for filtering alert search results by alert metadata, \textit{C. Query Results}: result set of alerts, and \textit{D. Geo-located Alerts}: a map for visualizing alerts' geo-location.}
	\label{fig:meco_overview}
\end{figure}

\meco\ includes a pipeline of three Stages, as depicted in Figure~\ref{fig:meco_overview}, whose respective goals are described as follows:
\begin{enumerate}
	\item \emph{Stage~I}. Identify, within the massive amount of daily tweets, those that are health related;
	\item \emph{Stage~II}. Create and monitor time series for each considered disease, looking for sudden peaks in the number of tweets, which could be an indicator of an outbreak;
	\item \emph{Stage~III}. Rank the potential tweets regarding an outbreak so that a public health official can manage the information associated to the event.
\end{enumerate}

The rest of the paper details these stages.

\section{Stage I: Message Filtering}
\label{sec:stage1}
A health-related term mentioned in a tweet can refer to many contexts, many of them not useful for the purposes of EI. For instance, tweets about vaccine, marketing campaigns, or ironic/jokes are non-relevant for an EI system. 

In essence, Stage~I's goal is to filter relevant health-related tweets from non-relevant ones. Moreover, the main novel idea behind the Stage~I is that, in presence of new outbreaks, the set and distribution of health-related terms used on Twitter should quickly change and consequently, we need a mechanism to handle this natural language changing within the Twitter stream. To the best of our knowledge, none of the works exploring the potential of Twitter for EI considers this idea of an evolving language.
%
%
%
%

\subsection{Challenges} 
Online message classification continues to be a complex and challenging task for long term EI surveillance and intelligence gathering in general. One reason for this is that given the evolution of real-world events, the
variable to observe cannot always be known a priori. One such example is
in the detection of food-borne illness, in which the contaminated food item is not known in advance. 

More in detail, the main challenges faced in Stage~I are as follows.

\textbf{Feature Change Detection.}~ 
Detecting new \textit{relevant} terms and filtering irrelevant tweets, are 
inter-related and impact each other. We need a way to: 1)~dynamically detect when \textit{new} and \textit{relevant}
terms appear in health-related tweets over time; and then 2)~subsequently incorporate the tweets containing these terms as part of the data to train our classification models. 

\textbf{Dynamic Labeling.}~As terminology evolves, the criteria for defining
a relevant tweet is likely to also change. However, expert labeling of
classifier training instances is expensive and in practice difficult to obtain, especially for the rate and volume needed to build and maintain a good
classifier.

In the rest of the section, we present the approach taken in \meco\ to address these challenges.

\subsection{Approach and Rationale} 
\label{sec:approach3}
%

The architecture of \meco's module encapsulating Stage~I has been designed around the idea of \textit{novelty management}. In particular, we incorporate a feature change detection mechanism into a framework for the adaptive classification of tweets.

Existing message classification techniques rely usually on supervised or unsupervised machine learning approaches~\cite{Fisichella2010,ICWSM112880}. In \meco, we defined a solution based on a semi-supervised approach to determine if tweets are relevant or not for outbreak detection and threat assessment. 

The choice of features and classifier was driven largely by the fact that the 
\meco\ system runs continuously, so features and classifiers that are not time-consuming to extract or encode need to be used. In the course of our investigation, we found that a linear Support Vector Machine (SVM) model exhibits a favorable trade-off between classification performance and training time. 

Our approach builds upon~\cite{Hido2008} and includes three main novel contributions beyond the state-of-the-art by 1)~modelling and describing the \emph{feature change} over time using an orthogonal vector, which is learned by a SVM; 2)~\mbox{computing} a
novelty score that lets the system identify those tweets that contribute to the
feature change, so that; 3)~manual labels can be obtained, dynamically and \mbox{on-demand}, by asking a human judge as part of an \textit{active learning} setting~\cite{Settles10activelearning}. 

One aspect that we assess in the experiments for this stage, is to what extent the expertise of the annotators impact the classification quality. To this end, besides public health experts, we also consider crowd-sourced workers from the CrowdFlower platform\footnote{\url{http://crowdflower.com/}} as annotators.

\begin{figure} [!tb]
	\centering
		\fbox{
\includegraphics[height=0.18\textheight]{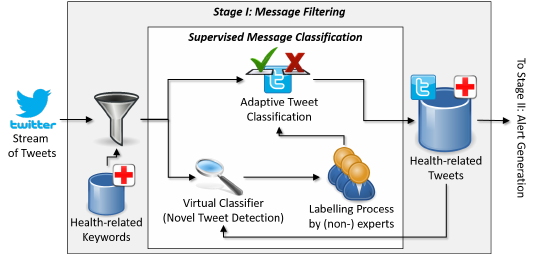}
		}
	\caption{Overview of our \meco\ approach for feature change detection
and adaptive classification.}
	\label{fig:stage1-overview}
\end{figure}

Figure~\ref{fig:stage1-overview} presents an overview of \meco's feature change detection and adaptive classification steps. The \texttt{Message Filtering} process runs continuously to classify all unlabeled tweets, which are arriving from the Twitter Streaming APIs. 

All tweets are annotated with locations, medical conditions, and temporal expressions using a series of language processing tools, including OpenNLP\footnote{\url{http://opennlp.apache.org/}} for tokenization, sentence splitting and part-of-speech tagging; HeidelTime for temporal expression extraction~\cite{StrotgenGertz2010:SemEval}; LingPipe dictionary-based entity extraction for medical conditions~\cite{lingpipe}; as well as entity extraction tools developed within the project for location entities~\cite{Lubomir}. In this paper, we only focus on \textit{text-based analysis} of Twitter messages. 

Then, we use two types of classifiers: a \texttt{Virtual Classifier (VC)} and an \texttt{Adaptive Classifier}. 

Positive keywords associated to diseases are pathogen (e.g., Streptococcus pyogenes) and symptoms (e.g., sore throat, fever, bright red tongue with a strawberry appearance, rash, bumps, itchy, and red streaks).

For filtering tweets irrelevant to the medical domain, we use a list of negative keywords associated to diseases from two freely-available resources:
1)~MedISys\footnote{http://medusa.jrc.it/medisys/homeedition/en/home.html} providing a list of negative keywords created by medical experts, and 2)~Urban Dictionary\footnote{http://www.urbandictionary.com/}, a Web-based dictionary of slang, ethnic culture words or phrases. 

The \texttt{Novel Tweet Detection} process is run periodically to trigger an update on the current adaptive classification model. In our case, a period of one week was used to correspond with weekly reporting performed by the health agencies in our study. 

The \texttt{Novel tweet Detection} phase is responsible to detect a feature change. To this end, the \texttt{Virtual Classifier} is used to compare incoming and unlabeled tweets against the set of existing tweets (those previously labeled during the \texttt{Adaptive Classification} step). If it is determined that feature change has occurred, novel tweets --~those for which it is expected that the classifier will not be able to correctly label~-- are flagged and channeled to the \texttt{Labeling Process} so that manual labels can be obtained by a human judge. 
%
%

Note that, regardless of whether feature change has been detected, all tweets are classified with a model that has been trained on the labeled instances that the \texttt{Adaptive Classifier} put into the \texttt{Health-related Tweets} storage. We apply the adaptive classification algorithm to all tweets presented to the system, even if feature change was detected, because the relevant tweets are needed by downstream components of our pipeline.  

The labels for novel tweets are acquired using \emph{dynamic labeling}, in which for-hire \textit{Human Intelligence Task} (HIT) or crowdsourcing, is used to obtain labels on-demand. Until the system receives HIT labels for the novel tweets, it uses temporary (automatic) labels given by the current \texttt{Adaptive Classification} model. 

After labeled tweets have been obtained from the HIT workers, they are used as input to the \texttt{Adaptive Classifier}; a new classification model is trained; and any temporary tweet labels are updated for improving and maintaining its accuracy over time.  Unlike other work~\cite{Paul2011,Demartini2012}, we also address the challenge of assessing the quality of the HIT labels against those of our domain experts (cf. Section \ref{sec:results}).
%
%
%
%

The VC detects feature change through a scoring mechanism in which the most novel tweets in the data stream are identified and presented to a human for labeling. Our experiments (cf. Section~\ref{sec:results}) show that the selected novel messages reflect the events in the real world that cause feature change and that they are useful for training and maintaining an accurate classifier for EI. 
%
 
The VC corresponds to a SVM model that maps a feature vector to a hypothetical class (i.e., novel/not-novel). The VC is trained to learn a decision boundary between the unlabeled incoming tweets and those existing in the collection. If this decision boundary is highly accurate, then it is very probable that feature change has occurred between the two data sets~\cite{Hido2008}.

\textbf{Novel Tweets Detection}. \meco\ assigns to each newly incoming tweet a score that represents how \emph{novel} it is to the system. This score is given with respect to the distance to the separating hyperplane of the aforementioned (binary) VC, which is trained on the set of labeled tweets and the set of new incoming tweets. We assume that the greater the distance between the newly incoming tweet and the tweets known to the system, the more dissimilar they are. 

Based on active learning principles, that is to label those instances that would most change the current model~\cite{Settles10activelearning,Georgescu:2014:DAC:2611040.2611047}, we propose that obtaining labels for the most novel tweets will help the system to keep its accuracy over time, a hypothesis that we test in our experiments detailed in Section~\ref{sec:results}.

\textbf{Labeling Criteria.} It is very challenging to determine if a tweet is relevant or not for EI, even for human experts. This is due to the fact that contextual information within the short messages is very limited, unlike existing work in the domain of EI based on news articles~\cite{2010_collier}.

In an effort to address the needs of our system for online and dynamic labeling of tweets, together with a team of
epidemiologist, we created a set of simple annotation guidelines for determining the relevance of tweets for the task: a tweet is \textit{relevant} if somebody reports himself or another person being ill while it is \textit{irrelevant} if no one is suffering from symptoms; i.e., mentions refers to opinion, advertising, jokes, music, books, films, artists, landmarks, sporting events, slang, etc.; and for the given set of criteria, we examine crowdsourcing as an alternative to obtaining the correct labels of tweets by the experts themselves. 

In the rest of the section, we present the results of the empirical evaluation to assess this approach.

\subsection{Results}
\label{sec:results}
The evaluation is composed of two separate experiments. In the first one, we were interested in evaluating the combined effectiveness of the Adaptive Classifier and of the Virtual Classifier over the time. 

In the second experiment, we evaluated the quality of the labelling process carried out by crowdsourcing compared with the one obtained by public health experts. 

\textbf{Feature change detection}.~We design the experiments to evaluate the suitability of the proposed approach to determine if tweets are relevant for outbreak detection as follows. 

We train and measure the performance of the Adaptive Classifier in a given time span and then use the model for prediction over a subsequent time span, with three different strategies to account for feature change:
\begin{enumerate}
	\item \textbf{No feature change handling} strategy in which a non-adaptive classifier and no retraining are used;
	
	\item \textbf{Random selection} of tweets used to retrain the Adaptive Classifier,
	
	\item \textbf{Novelty selection}, which uses the novelty scoring to select the tweets to retrain the Adaptive Classifier.
\end{enumerate}

In detail, we conduct the evaluations using a dataset consisting of 6,625 tweets collected within a time period from week 12th through week 14th of year 2011. These tweets were randomly selected from those containing at least one word in the \texttt{Health-related tweets} database (cf. Figure~\ref{fig:stage1-overview}). Then, we manually labeled them to determine if they were in fact health related or not, based on the guidelines outlined earlier in this section. Finally, we trained a SVM binary classifier using feature vectors derived from a bag-of-words representation combined with bi-grams.

We assess our results in terms of \textit{accuracy}, i.e., the proportion of true results (both true positives ($TP$) and true negatives ($TN$) among the total number of tweets examined:
\begin{equation*}
\label{eq:accuracy}
accuracy := \frac{TP+TN}{\text{Total Number of Tweets}}
\end{equation*}
Here, $TP$ corresponds to the number of tweets classified and labeled as health-related, and $TN$ to the number of tweets classified and labeled as non health-related. 

We then applied this classifier to the feature change detection task. We employed another dataset of 6,625 tweets, now randomly selected from the calendar weeks 15th through 19th of year 2011. We created a subset of about 1,100 positive and negative tweets for each week and manually labelled them as well.
%

\begin{figure}[t]
     \centering   
           \subfloat[Accuracy, $q=1\%$]{%
           \fbox{
              \resizebox {0.3\textwidth} {!} {
	\begin{tikzpicture}
			\begin{axis}[ylabel=Accuracy, xlabel=Time Slot (Week),ymin=75,ymax=95,
			cycle list name=bw]
			\addplot coordinates {
				(15,85.53)
				(16,85.34)
				(17,81.23)
				(18,84.03)
				(19,84.33)
			};
\addlegendentry{No Feature Change Handling}
			\addplot coordinates {
				(15,85.62)
				(16,85.62)
				(17,81.14)
				(18,83.85)
				(19,84.98)
			};
\addlegendentry{Random Selection}
			\addplot coordinates {
				(15,85.43)
				(16,85.62)
				(17,83.19)
				(18,84.69)
				(19,84.70)
			};
						
\addlegendentry{Novelty Selection}
			\end{axis}
		\end{tikzpicture}

}
}
              \label{fig:oscHigh}%
           } 
           \subfloat[Accuracy, $q=5\%$]{%
           \fbox{
             \resizebox {0.3\textwidth} {!} {
	\begin{tikzpicture}
			\begin{axis}[ylabel=Accuracy, xlabel=Time Slot (Week),ymin=75,ymax=95,
			cycle list name=bw
			]
			\addplot coordinates {
				(15,85.53)
				(16,85.34)
				(17,81.23)
				(18,84.03)
				(19,84.33)
			};
\addlegendentry{No Feature Change Handling}
			\addplot coordinates {
				(15,85.71)
				(16,84.97)
				(17,83.47)
				(18,85.53)
				(19,85.26)
			};
\addlegendentry{Random Selection}
			\addplot coordinates {
				(15,85.81)
				(16,85.43)
				(17,84.97)
				(18,84.87)
				(19,85.73)
		};
\addlegendentry{Novelty Selection}
			\end{axis}
		\end{tikzpicture}

}}
              \label{fig:oscHigh1}%
           }
           \subfloat[Accuracy, $q=10\%$]{%
           \fbox{
           \resizebox {0.3\textwidth} {!} {
	\begin{tikzpicture}
			\begin{axis}[ylabel=Accuracy, xlabel=Time Slot (Week),ymin=75,ymax=95,
			cycle list name=bw]
			\addplot coordinates {
				(15,85.53)
				(16,85.34)
				(17,81.23)
				(18,84.03)
				(19,84.33)
			};
\addlegendentry{No Feature Change Handling}
			\addplot coordinates {
				(15,86.18)
				(16,84.78)
				(17,84.78)
				(18,84.59)
				(19,87.03)
			};
\addlegendentry{Random Selection}
			\addplot coordinates {
				(15,87.49)
				(16,85.81)
				(17,86.65)
				(18,84.78)
				(19,86.38)
			};
\addlegendentry{Novelty Selection}
			\end{axis}
		\end{tikzpicture}

}}
              \label{fig:oscHigh2}%
           }
\caption[Accuracy and feature change test statistic on the labeled English
Twitter data set]{Classifier accuracy during a weekly feature change detection phase.}
	\label{fig:EngTwitter}
	
\end{figure}
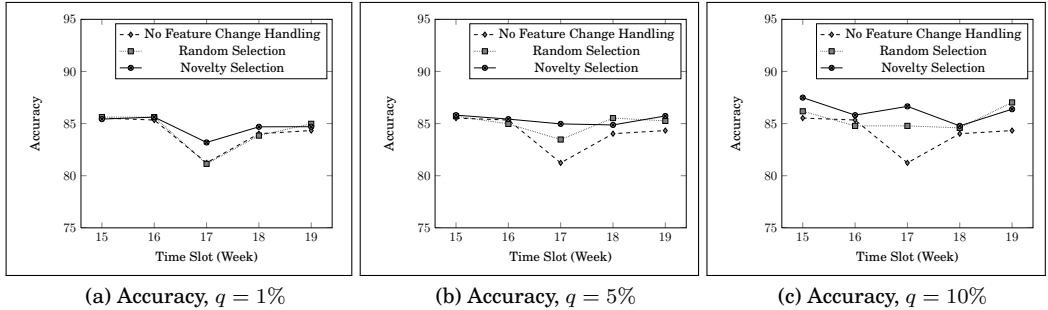

We considered three different scenarios with a varying percentage, $q$, of the novel tweets. In particular we evaluated the scenarios where $q=1\%$,  $q=5\%$ and $q=10\%$, using a significance level, $\alpha=0.01$. 
Results are reported in Figure ~\ref{fig:EngTwitter}. 

From these results, we can derive the following conclusions:
\begin{enumerate}
	\item The strategy using the Virtual Classifier provides on average the best results, and the more novel tweets are, the better is the accuracy of this strategy.
	
    \item The strategy without any change provides the poorest results, highlighting that the dictionary evolves over the time (even over the limited timeframe we considered) and some mechanism to update the classifier is needed.
    
	\item The accuracy of the classifiers degrades for most strategies during week~17th, 2011. This time slot contains many tweets mentioning ``royal wedding fever'', referring to the wedding of Prince William and Kate Middleton on April 29, 2011. Those tweets are sometimes misclassified as relevant. 
\end{enumerate}



%

\textbf{Crowdsourcing versus Experts.}~
The second experiment explores the quality of the \textit{labelling process} performed by human annotators. 
To this end, we employed a total of 1,500 tweets, by randomly sampling 500 tweets from each of the
calendar weeks 15th, 16th and 17th, 2011. We presented them to 43 workers of the CrowdFlower
platform to elicit their feedback. 
%

To control the quality of the crowdsourced labels, two actions were taken.
First, a set of ``golden'' tweets with known labels was added to the unlabeled
tweets that were randomly shown to the workers. A trust value is computed for
each worker based on the number of correctly labeled ``golden'' tweets. If this
trust value is below a fixed threshold, the workers labels are removed from the
task. Second, each tweet was labeled by a minimum of 3 workers for
each task, and only those tweets corresponding to a majority agreement above 65\% percent  
among the multiple annotators were used. 

The average agreement among the 43 HIT workers on a given tweet was 93.89\% with an average accuracy on the injected gold labeled tweets of 92\%.  

Out of the 1,500 tweets labeled by the crowd, 1,114 tweets had a perfect agreement of 100\%;  295 tweets
had an agreement between 66\% and 100\%; and 91 tweets had an agreement between
50\% and 66\%.

We chose 130 tweets according to the agreement of the crowdsourcing annotators to include a representative amount of ``easy'' tweets (perfect agreement) and more ``difficult'' tweets (low agreement). Then, we asked five public health experts to also annotate this subset and measured their inter-annotator agreement on the 130 instances, which achieved a 89.33\%.

For the 130 tweets labeled by both the experts and the HIT workers, there was a percent agreement of 87.69\%. The percent agreement is the ratio of number of agreements among crowd and experts to the total number of units judged by crowd and experts.

When measuring the classifier performance individually for each group, based on a 10-fold cross
validation and equal percentages of positive and negative examples, the classifier performance in terms of accuracy was 75\% for the experts and 83\% for the HIT workers. 

\subsection{Lessons Learned and Outlook}
\label{sec:outlook1}
Our results suggest that people outside the public health domain are able to accurately judge the relevance of tweets when given  a simple set of criteria. Thus, once the modules of Stage I have detected a feature change, it is also feasible to outsource the novel tweets as part of a separate feature change handling
procedure, without the necessary involvement of a public health expert.
%

Although the classifier performance was less for the expert labels, than
the crowd labeled data, we believe this is due to the fact that in practice, whether a tweet is relevant for an expert depends on several factors, such as different time periods of an outbreak (e.g., before, during or after); or on the task and role of the expert with respect to an epidemic investigation. Nonetheless, the crowd can still help to filter label instances that are clearly off topic.

To get a better understanding of the impact of detected feature change on the
classification accuracy, a larger set of expert labeled tweets for
experimentation would be useful to further improve the significance of the
results. However, by doing so, it would still not address the need for experts to
re-label each time feature change is detected and in practice, the overhead of
such a task is too expensive and not timely enough. We propose instead, that 
after tuning expert labeled examples with a good inter-annotator agreement, such instances could be used as gold standard to filter out HIT workers whose trust value is below a threshold.

\section{Stage II: Alert Generation}
\label{sec:stage2}
In this section, we describe key challenges faced during the alert generation process, we compare different biosurveillance algorithms, and provide experimental results as well as a discussion on the impact of the identified challenges in this stage.

\subsection{Challenges} 
\label{sec:challenges2}
The majority of existing works center their studies on recurring diseases only, e.g., Influenza-like Illness (cf. Section~\ref{sec:realted}). Moreover, many researches are either focused on a single nation and language, or the spatial dimension is not considered at all.

\begin{table}[!tb]
\tbl{The 5 considered outbreaks, with ID, disease (or medical condition), country, and duration of the event.}{
	\centering
	\begin{tabular}{clll}
	\toprule
	\multicolumn{1}{c}{\textbf{ID}} & \multicolumn{1}{l}{\textbf{Disease}} & \multicolumn{1}{l}{\textbf{Country}} & \multicolumn{1}{l}{\textbf{Event period in 2011}}  \\ 
	\midrule
 1 & Anthrax & Bangladesh & [June -- August]  \\ 
 2 & Botulism & France  & September   \\ 
 3 & Cholera & Kenya & [November -- December]  \\ 
 4 & Escherichia Coli & Germany & [May -- July]  \\ 
 5 & Mumps & Canada & [June -- August] \\ 
	\bottomrule
	\end{tabular}
	}
	\label{tab:OutbreakList}
\end{table}

Our goal is to provide a broader EI system, able to detect and monitor outbreak events in Twitter, for multiple locations and for multiple diseases, including sudden and unexpected outbreaks (cf. Table~\ref{tab:OutbreakList}). This goal is a challenging task, due to two main issues: 

\textbf{Spatio-Temporal Monitoring of Diseases.}~Location-awareness is one of the key starting points for any EI solution. Indeed, knowing where an outbreak is happening is naturally one of the most important pieces of information. 

From this point of view, the typical use of Twitter is not helping. For instance in our experimentation, we observed that explicit coordinates where present in less than 1\% of the collected tweets. As a consequence, other techniques, mainly based on Natural Language Processing, or on the analysis of user profiles of people tweeting are needed in order to infer the location of a tweet.

\textbf{Temporal and Spatial Dynamics of Diseases.}~The characteristics of infectious diseases are highly dynamic in time and space, and their behavior varies greatly among different regions and the time periods of the year. E.g., some infectious diseases can be rare or aperiodic, while others occur more periodically. In addition, various diseases have different transmission rates and levels of prevalence within a region. 

To get a deeper insight on this issue, we collected information about five outbreaks that occurred in 2011, which are detailed in Table~\ref{tab:OutbreakList}.

By analyzing the time series of the tweets about these five different outbreaks, we found out that the Twitter data regarding the public health outbreaks can be characterized by two dimensions: (1)~\emph{Oscillation}, which is seen as the frequency at which the curve spikes, and (2)~\emph{Magnitude} (or volume) of daily count of tweets, sinks or slopes.

\begin{figure}[!t]
	\begin{centering}
	{
	\includegraphics[width=0.8\linewidth]{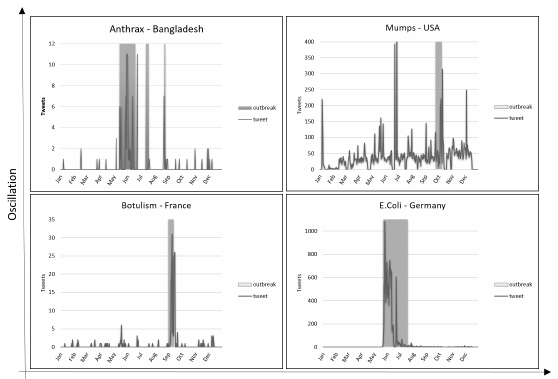}\\
	{\scriptsize Magnitude}
	}
	\caption{Different temporal dynamics of outbreak-related tweets based on their \emph{oscillation} and \emph{magnitude}.}
	\label{fig:oscillations}
	\end{centering}
\end{figure}

Figure~\ref{fig:oscillations} shows representative examples of the different outbreak dynamics from Twitter time series data. The gray areas in these plots represent the timeframe where an outbreak alert was broadcasted via \promed~\cite{promed}, a global reporting system providing information about outbreaks of infectious diseases.

A time series with low oscillation indicates that the average daily number of tweets is more or less constant (eventually zero) but it noticeably peaks within the outbreaks period.
Examples are provided by the two bottom frames of Figure~\ref{fig:oscillations}. On the left there is the distribution of tweets for the outbreak of Botulism in France in 2011. We can see that the magnitude of the tweets is very low, both outside and inside the emergency time frame. This situation may occur in the scenarios where the diffusion of English tweets is limited. 

On the other hand, the picture on the right is the case of the Escherichia Coli outbreak in Germany, 2011. Since it had a very high international and media coverage, due to the ease with which it could spread, we can see that the number of tweets during the peak is almost two orders of magnitude higher than the example of Botulism in France. 
In general, in presence of diseases leading to low oscillations of tweets, it is easy for surveillance algorithms to produce correct alerts.

A time series with a high oscillation means that the number of tweets varies often and greatly over the year. Considering the outbreak Anthrax in Bangladesh (Figure~\ref{fig:oscillations}, upper left frame) as an example of high oscillation/low magnitude time series, we can observe a number of tweets per day ranging from 0 to 5 all over the year. Thus, a less sensitive surveillance algorithm is necessary to avoid to get continuously false positives. 

A time series with high oscillation and high magnitude occurs when:  1)~a disease occurs continuously in a country, such as Mumps or Leptospirosis and/or 2)~the name of the disease is a highly ambiguous term, such as for Mumps, which is for instance also the name of a software system\footnote{MUMPS is also the acronym for \textit{Massachusetts General Hospital Utility Multi-Programming System} -- \url{https://en.wikipedia.org/wiki/MUMPS}}.

In Figure \ref{fig:oscillations}, upper right frame, we report on the Mumps outbreak in the USA, which is not included in the evaluation since it was not possible to find a reliable ground-truth regarding the dates of the emergency.
%

In these situations, algorithm tuning is essential, since the dynamic of data is too large, and it is not easy to identify significant aberrations. 

\subsection{Approach and Rationale}
\label{sect:Appr2}

The Stage~II of our EI system takes as input the dataset containing health-related tweets coming from the Stage I, and provides two outputs: (1)~a collection of datasets containing health-related tweets, aggregated by the spatial dimension and (2)~a set of \textit{alerts}, triggered by surveillance algorithms analyzing the temporal dynamics of each of these datasets. 

Thus, Stage~II encompasses two main steps, \textit{context creation} and \textit{biosurveillance}, which are described as follows.

\textbf{Context creation}. We are interested in monitoring, for each \emph{location}, a set of possible diseases. In order to infer the location associated to a tweet we define three rules, which are applied in order of importance, as follows:
%
\begin{enumerate}
	\item Mention of the location (city, country, etc.) in the text of the tweet.
	
	\item The tweet's geo-location information (latitude and longitude), if present. 
	
	\item Location indicated in the user profile of the author of the tweet.
\end{enumerate}

Tweets not matching any of the above rules are discarded. Even if they are health-related, it is not possible to understand where the event is located. Thus, their informative contribution is negligible. Moreover, we do not consider the tweet language (provided as an attribute by the Twitter API) in determining location information, since it is too inaccurate.

In the end, the geographical granularity level we considered is the country. The geo-mapping between location or coordinates and the relative nation was performed using the Yahoo BOSS Geo Services APIs \footnote{https://developer.yahoo.com/boss/geo/}.

The output of this step is a collection of datasets, which are intended as a spatial partitioning of the original dataset coming from the Stage I. The total sum of contained tweets is most likely lower, since tweets without any information on the location are discarded.
%
%
%

\textbf{Biosurveillance}. A standard approach to detect anomaly in health-related time series data is to leverage state-of-the-art biosurveillance algorithms. Many of them can be found in the literature~\cite{farrington_1996,Hutwagner2003,DBLP:journals/jbi/Khan07} and implemented in analysis tools like R~\cite{R:surveillance}. 

In our study, all biosurveillance algorithms are included in the free package \textit{Surveillance} for R, that implements multiple statistical methods for the ``Temporal and Spatio-Temporal Modeling and Monitoring of Epidemic Phenomena''~\cite{R:surveillance}.

In particular, we assessed the following four algorithms: C1, C2, C3, and Farrington, which are described as follows.

	\noindent\emph{Early Aberration Reporting System} (EARS), which compute a test statistic on day $t$ as follows: $S_t = \max(0, (X_t - (\mu_t  + k \, \sigma_t))/\sigma_t)$, 	where $X_t$ is the count of episodes on day $t$, $k$ is the shift from the mean to be detected, and $\mu_t$ and $\sigma_t$ are the mean and standard deviation of the counts during the baseline period.
	EARS uses three baseline aberration detection methods, that we assessed:
	\begin{itemize}
		\item \texttt{C1-Mild}, where the baseline is determined on the average count from the past 7 days;
		\item \texttt{C2-Medium}, where the baseline is determined on the average count from the 7 days in the 10 days prior to 3 days prior to measurement.
		\item \texttt{C3-High}, that uses the same baseline as C2, but takes a 3 day average of events to determine the measure.
	\end{itemize}
	
	\noindent The \emph{Farrington (FA)} detection algorithm predicts the observed number of counts based on a subset of the historic data, by extracting reference values close to the week under investigation and from previous years, if any. The algorithm fits an overdispersed Poisson generalized linear model with log-link to the reference values.~\cite{farrington_1996} 
	
	The rationale behind the selection of these methods is that the EARS family require a very limited number of previous data to provide an alert, thus being potentially suitable for a new outbreak detection system. Moreover, it has basically no parameters to tune, make its applicability straightforward.
	
	On the other hand, the Farrington algorithm is largely considered a robust and fast method. Thanks to these characteristics,  currently it is the method used at European public health institutes~\cite{hulth2010practical}. This algorithm has a set of parameters to specify. In particular, among others, it requires to define the windows size, $w$, i.e. number of weeks to be considered for the alert generation. We performed our experimentations with $w$ = 2, 3, and 4. 
	Nevertheless, since the choice of the parameters heavily depends upon the data, it is outside the scope of this research to investigate hyperparameter optimization techniques (e.g. as in~\cite{DiMartinoEsem13}).
	
	Further details on these algorithms and their implementation can be found in the R-Surveillance documentation~\cite{hohle2013surveillance}.
%
%
%
%

\subsection{Results}
Here we seek to address the following question: \textit{what are the most suitable surveillance algorithms for outbreak alert generation using Twitter data?}
									
To perform our study, we analyze Twitter data collected from January the 1st, 2011 to December the 31st, 2011. The data was collected using the pipeline defined in the Stage~I (Section~\ref{sec:stage1}), resulting in a total of 112,134,136 health-related tweets.

\textbf{Ground Truth.}
Studying the usefulness of Twitter data in an early warning task requires real-world outbreak statistics. Therefore, we build a ground truth by relying upon \promed~\cite{promed}. An outbreak event is intended as a temporal anomaly found in time series data that occur when the impact of an infectious disease is above an expected level at a certain time.

We collected 3,056 \promed\ reports occurred during year 2011, and among them, we selected 5 different outbreaks according to two criteria: 1) a clearly identifiable starting date from the \promed, by considering the first \promed\ post on it, and 2) a representative distribution of tweets. 

An important aspect of our work is that we consider the duration of each outbreak by manually analyzing the text of each \promed\ document, unlike previous work~\cite{2010_collier} that assumes the publication date of a document as the estimated relevant time of an outbreak. 

In particular, we determine the starting date of a disease by looking at the text inside the first \promed\ post, and the ending date was associated to the text inside the last \promed\ publication, for that particular disease-location pair. One reason for doing this is that the events in \promed\ undergo moderation, so there is often a delay between the time of the actual outbreak and the publication date of the related report. However, it is worth noting that this strategy gives us a good confidence only on the beginning date of the outbreak. In fact, the absence of further \promed\ posts does not necessarily mean an end of the outbreak, but just that there was no significant news in which it was reported.

\textbf{Evaluation Metrics.}
To assess the quality of the generated alerts, we use standard Information Retrieval metrics, namely \emph{precision}, \emph{recall}, and \emph{f-measure}, which are defined as follows:
%
\[
precision := \frac{TP}{TP + FP} \;;\; recall := \frac{TP}{TP + FN} \;;\; f\text{-}measure := 2 * \frac{precision * recall}{precision + recall} \; .
\]
%
%
%

With our problem at the hand, a valid definition of the temporal granularity is not easy, due to the high temporal variability of the data and the lack of a proper ground truth. To clarify it, in case an algorithm generates an alert 5 days before the official communication in \promed, should it be considered a false positive or a true positive with a good timeliness? We opted for the second case, thus defining a $TP$ \emph{(True Positive)} as an alarm that is raised within the time frame of the \promed alert or up to 10 days before it. 
A $FP$ \emph{(False Positive)} is an alert generated outside this above defined time frame. A $FN$ \emph{(False Negative)} is an alert not generate in this time frame.

\begin{figure}[!tb]
	\centering
	\subfloat[Anthrax in Bangladesh]{\label{fig:ResultsLO:anthrax_bangladesh}\includegraphics[width=0.49\textwidth]{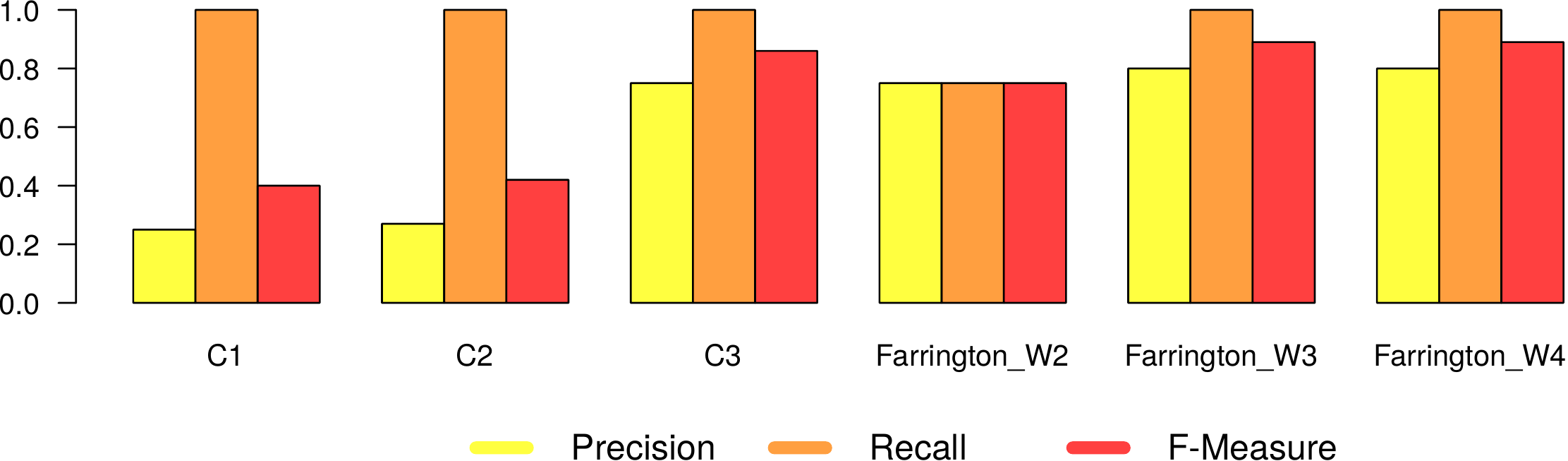}}
	~
	\subfloat[E. Choli in Germany]{\label{fig:ResultsLO:ecoli_germany}\includegraphics[width=0.49\textwidth]{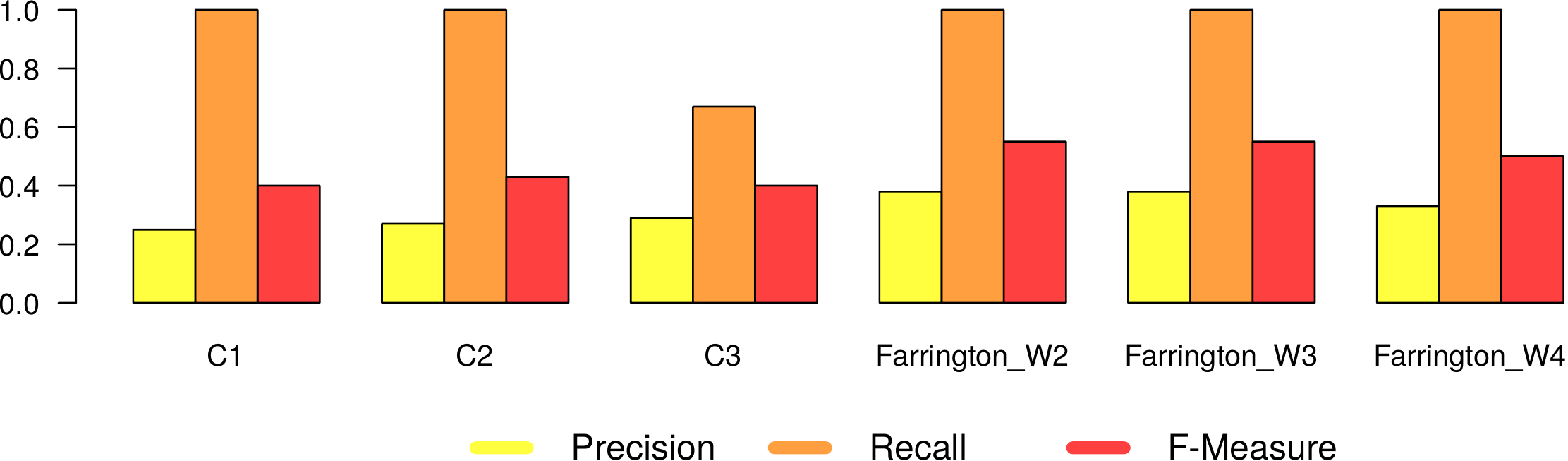}}
	\caption{Performance of the investigated algorithms on the diseases with low oscillation.}
	\label{fig:ResultsLO}
\end{figure}

\begin{figure}[!tb]
	\centering
	\subfloat[Botulism in France]{\label{fig:ResultsHO:botulism_france}\includegraphics[width=0.49\textwidth]{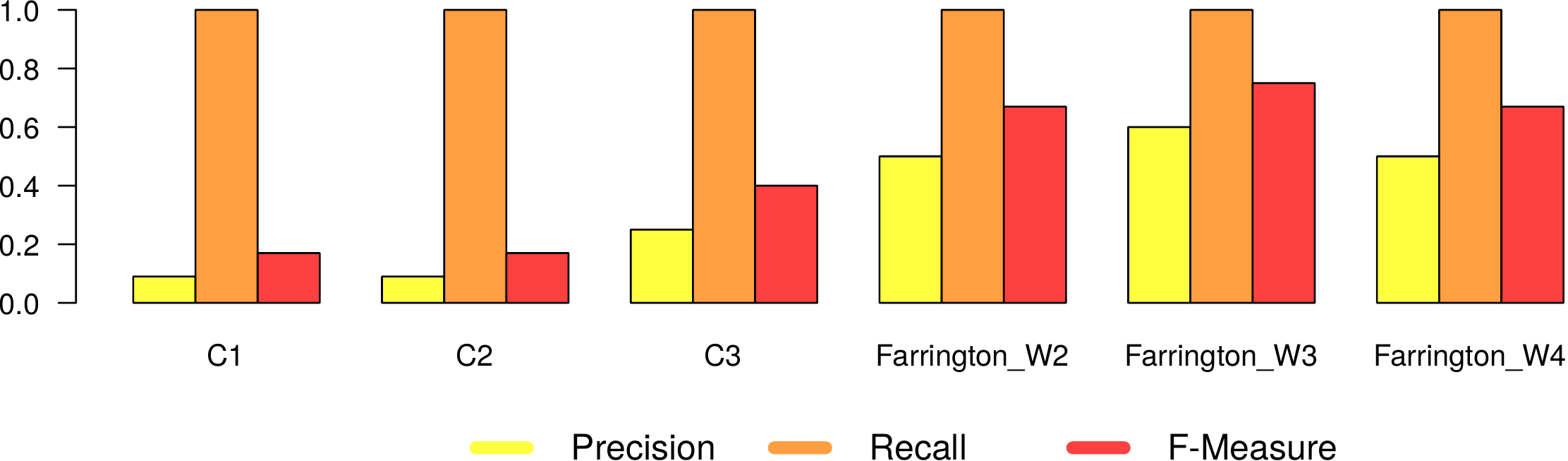}}
	\\
	\subfloat[Cholera in Kenya]{\label{fig:ResultsHO:cholera_kenya}\includegraphics[width=0.49\textwidth]{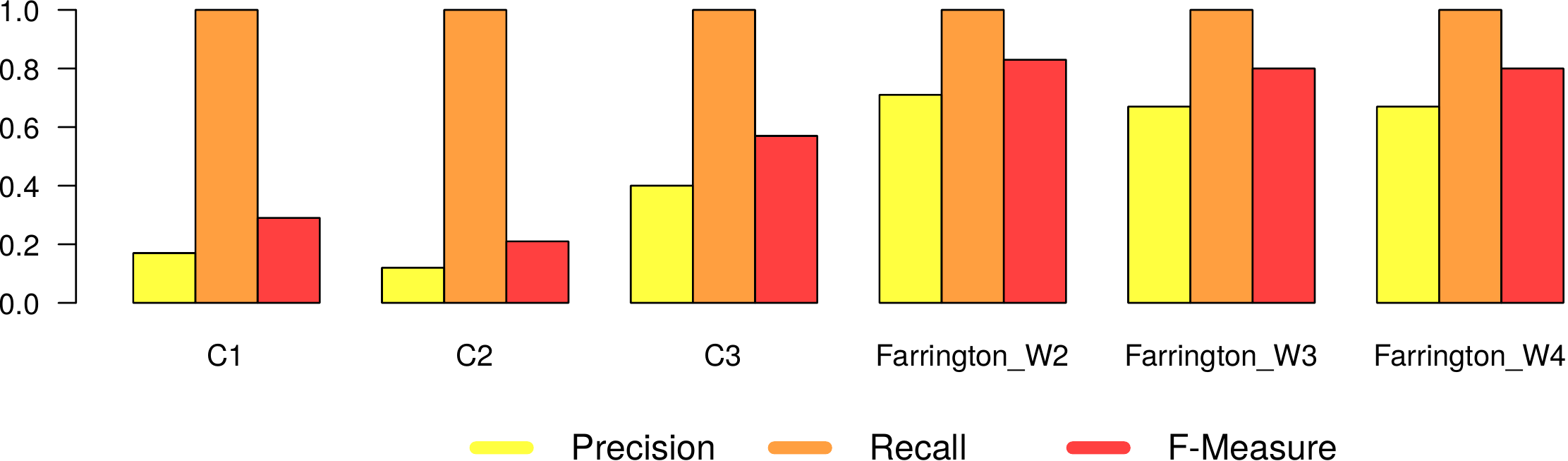}}
	~
	\subfloat[Mumps in Canada]{\label{fig:ResultsHO:mumps_canada}\includegraphics[width=0.49\textwidth]{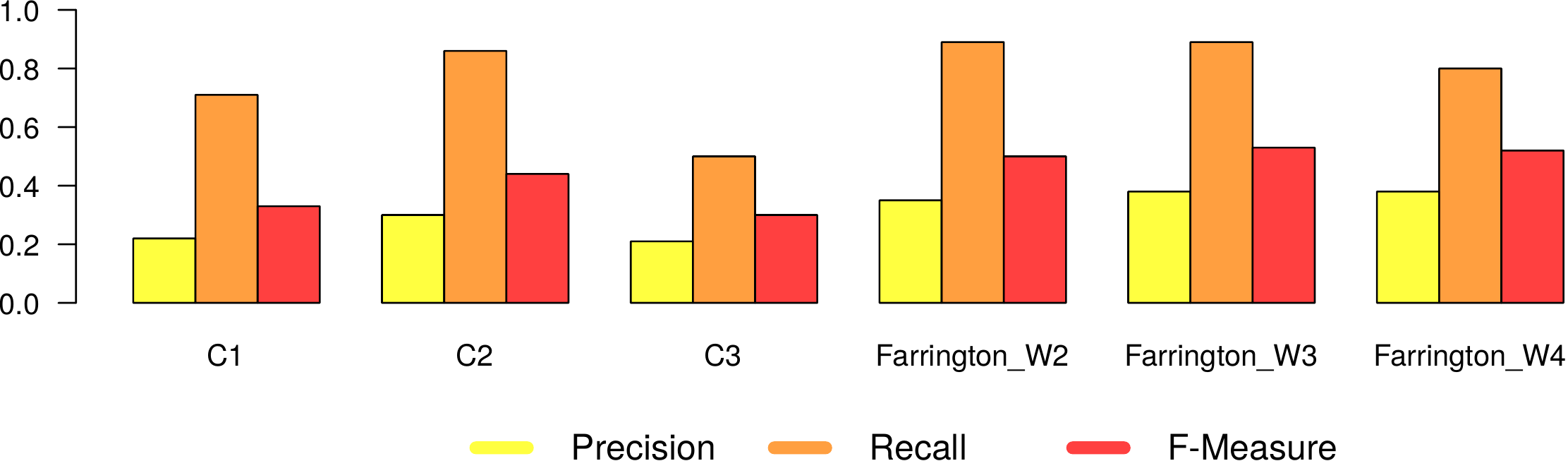}}
	\caption{Performance of the investigated algorithms on the diseases with high oscillation.}
	\label{fig:ResultsHO}
\end{figure}

In Figures \ref{fig:ResultsLO} and \ref{fig:ResultsHO}, we present the results we got from the application of the 4 different surveillance algorithms to our dataset. 
In particular, as described before, for the three EARS algorithms, we used the default parameters, while for the Farrington algorithm, we tested the values 2, 3 and 4 as for the windows size. Consequently in Figures \ref{fig:ResultsLO} and \ref{fig:ResultsHO} we report 6 different results (namely EARS C1, C2 and C3, Farrington W2, W3 and W4) for the five evaluated diseases.

%
%

\subsection{Lessons Learned and Outlook}
\label{sec:lessonsTw}

As mentioned earlier, our goal is to detect outbreak events for general diseases that are not only seasonal, but also non-recurring diseases. From the results we collected, we can report the following lessons learned:

	--~In general all the algorithms present a very high recall (very close to 1), meaning that they are able to detect very well the epidemic outbreaks. This is an expected result, given the high correlation that is clearly visible in our datasets between tweet peaks and outbreak timeframes. 
	
	\noindent The only notable exception comes from Mumps in Canada, which can be classified as high oscillation -- high magnitude type of outbreak. 
	
	--~The real difference among the algorithms can be seen in terms of number of false alarms (i.e., precision). We can notice that in general the EARS family is providing a large amount of false positives. Probably this is due to the short time window used from these algorithms. On the other hand, Farrington algorithm performs better in terms of precision and f-measure, regardless of the window size. Thus we can in general confirm the effectiveness of the Farrington algorithm for this kind of task. As already reported in Section \ref{sect:Appr2}, this is the standard algorithm for epidemic outbreaks in many health institutions. In our case, the Farrington results are always better than those of the three EARS algorithms. 
	
	--~The hyperparameter \textit{window size} of the Farrington algorithm has an impact on the quality of the provided alarms, but no generalizable trend can be devised from our experiments. The window size should be defined according to the specific temporal dynamics of each considered disease. A solution to automatically find the best hyperparameter optimization would be highly recommendable to solve this issue.
	
	--~As expected, there is a noticeable difference between low and high oscillation diseases. In the latter case, all the algorithms perform significantly worse.
\section{Stage III: Recommendation Techniques for Threat Assessment}
\label{sec:stage3}
After the detection of the outbreak  in \meco's Stage II, authorities investigating the cause and the impact in the population are interested in the analysis of tweets related to the event. However, thousands of messages could be produced every day for a major outbreak, which make such task overwhelming for the investigators who are quickly inundated with the volume of tweets that must be examined when assessing threats. 

The goal of Stage III is to facilitate access for the end user to the original tweets, organized by the alerts, and ranked based on his or her interest for the task of outbreak analysis and control. In \meco, we employed recommendation techniques to tackle this problem. In this section, we present the particular challenges we face in this stage and review one of the recommendation methods used within the system, namely \textit{Personalized Tweet Ranking for Epidemic Intelligence} (PTR4EI)~\cite{ICWSMDIAZ2012,WWWDIAZ2012}.
%
%

\subsection{Challenges}
Even though algorithms for recommender systems and learning to rank are agnostic to the problem domain, their application for EI based on Twitter is not straightforward. We identified two major challenges faced within \meco:

\textbf{Limited User Feedback Available for the Recommender System.} Learning to Rank and Recommender Systems approaches have been successfully
applied to address the growing problem of information overload in a broad range of domains, for instance, web search, music, news media, movies, and collaborative annotation~\cite{recsys_handbook_2011}. Such approaches usually build models offline, in a \textit{batch} mode, and rely upon abundant user interactions and/or the availability of explicit feedback (e.g., ratings, likes, dislikes). In the case of EI, experts' interactions and explicit feedback are scarce, which makes it harder to build effective models for ranking or recommendation.

\textbf{Dynamic Nature of Twitter.} The real-time nature of Twitter, on the one hand, makes it attractive for public health surveillance; yet, on the other, the volume of tweets also makes it harder to: 1)~capture the information transmitted, 2)~compute sophisticated models on large pieces of the input, and 3)~store the input data, which can be significantly larger than the algorithm's available memory~\cite{muthu}. One major challenge in monitoring Twitter for EI lies in capturing the dynamics of an outgoing outbreak, without which the time-sensitive intelligence for threat assessment would be rendered useless.

 
\subsection{Approach and Rationale}
Our PTR4EI extends a learning to rank framework~\cite{Liu:2009:LRI:1618303.1618304} by considering a personalized setting that exploits a user's individual \textit{context}. We consider such context as implicit criteria for selecting tweets of potential relevance and for guiding the recommendation process. The user context $C_u$ is defined as a triple $C_u = (t, MC_u, L_u)$, where $t$ is a discrete \textit{Time} interval, $MC_u$ the set of \textit{Medical Conditions}, and  $L_u$ the set of \textit{Locations} of user interest.

We consider an initial context $C_u$ specified explicitly by the user, which can be precise, but static and limited to the medical conditions or locations manually included by the user. Our goal is to automatically capture the dynamics of the outbreak as reported in Twitter. To this end, we expand the user context by including additional medical conditions and locations related to the ones she specified, and exploit the resulting and richer context for personalized ranking. Our approach expands the user context by using 1)~latent topics computed with LDA~\cite{LDA} based on an indexed collection of tweets for epidemic intelligence; and 2) hash-tags that co-occur with the initial context. 

We use the terms in the expanded context that correspond to medical conditions, locations, and complementary context\footnote{\textbf{Complementary Context} corresponds to the set of nouns, which are neither Locations nor Medical Conditions. It may include named entities such as names of persons, organizations, affected organisms, expressions of time, quantities, etc.} to build a set of tweets by querying our collection, which correspond to a subset of Tweets output by Stage~II. This step helps us to filter irrelevant tweets for the user context. 

Next, we elicit judgments from experts on a subset of the tweets retrieved in order to build a ranking function model. We then obtain for each labeled tweet a feature vector that help us training our personalized ranking function. Finally, we use the ranking function to rank new incoming tweets automatically. Please refer to \cite{ICWSMDIAZ2012} and \cite{WWWDIAZ2012} for further details.

\subsection{Results}
In this section we review the experimental evaluation of our approach on the EHEC outbreak in Germany, 2011, as the real-world event of interest, and discuss the results we obtained.

\begin{figure}[!tb]
\centering
\includegraphics[height=0.15\textheight]{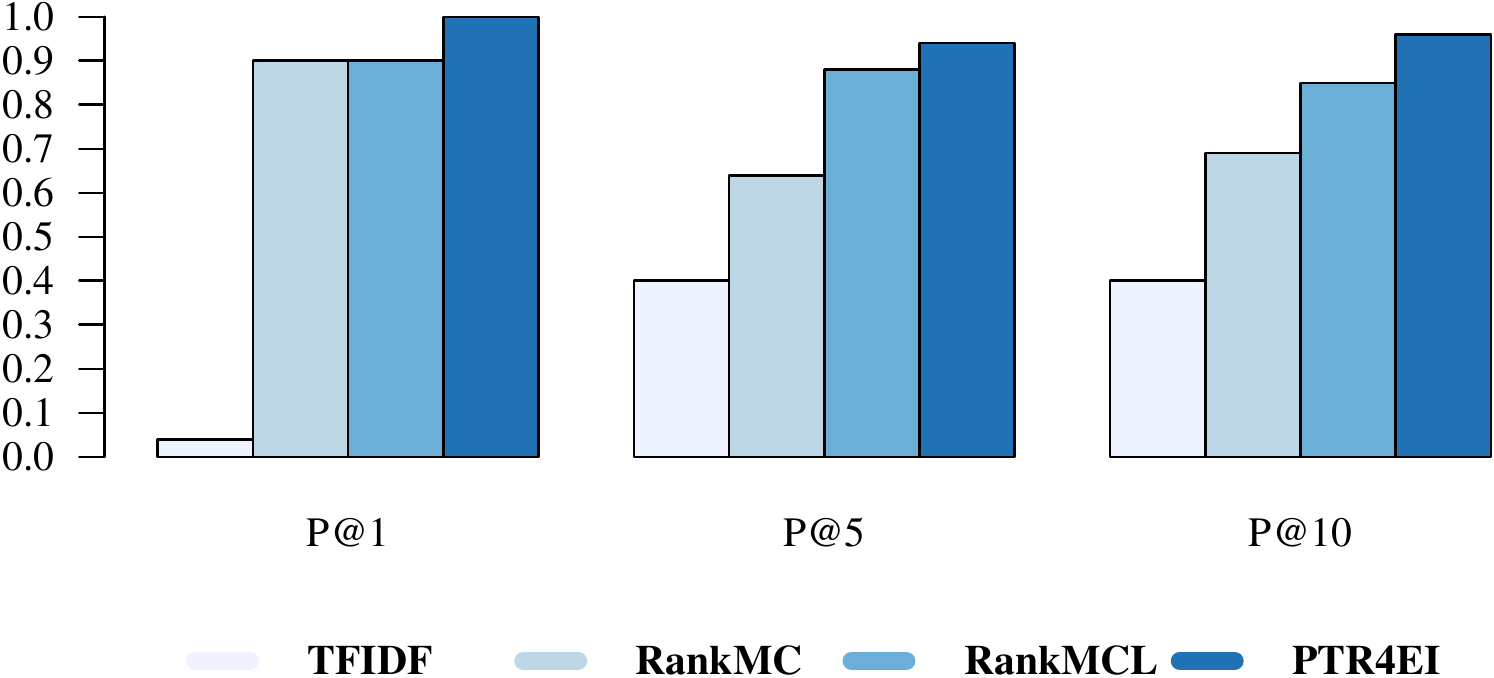}
\caption{\small Personalized ranking performance in terms of Precision@\{1, 5, 10\}.}
\label{fig:results_precision}
\end{figure}

To support users in the assessment and analysis during the German EHEC outbreak, we got an alert from Stage II starting on 2011-05-23. We monitored related tweets up to 2011-06-19. In this way, we are taking into account the main period of the outbreak,\footnote{Note that even though the main period of the outbreak is considered for the evaluation, nothing prevents us to build the model during the ongoing outbreak, and recompute it periodically (e.g., weekly).} the disease of interest, and the location. 

During the period of the EHEC outbreak (May and June, 2011) a total of 7,710,231 tweets related to medical conditions were collected by the \meco\ system (Stage~I); of which, 456,226 were related to the EHEC outbreak in Germany. Individual judgment were solicited to three experts on a subset of 240 of these tweets. The experts were asked to provide a relevance judgment as to whether the tweet was relevant in supporting their analysis of the outbreak, or not; disagreement in the assigned relevance scores were resolved by majority voting.

For each tweet, we prepared five binary features: $F_{MC}$, $F_{L}$, $F_{\text{\#-tag}}$, $F_{CC}$, and $F_{URL}$. We set the corresponding feature value equal to \textit{true} if a medical condition, location, hash-tag, complementary context term, or URL were present in the tweet, and \textit{false} otherwise. For learning the ranking function, we used the Stochastic Pairwise Descent algorithm~\cite{Sculley:2010}.

We compared our approach that expands the user context with latent topics and social generated hash-tags, against three ranking methods:

   \noindent\textbf{-- TFIDF} is a vector space model based on a truncated list of documents, which are retrieved from an indexed twitter collections using the conjunctive query: `` EHEC \texttt{AND} `Lower Saxony' " and sorted using TF-IDF scores.

    \noindent\textbf{-- RankMC} learns a ranking function using only medical conditions as feature, i.e., $F_{MC}$. Please note, that this baseline also considers additional medical conditions that are related to the ones in $MC_u$, which makes it stronger than non-learning approaches, such as BM25 or TF-IDF scores that use only the $MC_u$ elements as query terms.
    
    \noindent\textbf{-- RankMCL} is similar to RankMC, but besides the medical conditions, it uses a local context to perform the ranking (i.e., features: $F_{MC}$ and $F_{L}$).

We randomly split the dataset into 80\%  training tweets, which will be used to compute the ranking function, and 20\%  testing tweets. To reduce variability, we performed the experiment using a cross-validation with 10 different 80/20 partitions. The test set is used to evaluate the ranking methods. The reported performance is the average over the 10 rounds. For evaluation, we used  precision at position $n$ ($P@n$)~\cite{2011_ricBaeza}. 

The ranking performance in terms of precision is presented in Figure~\ref{fig:results_precision}.  As can be seen,  PTR4EI outperforms the three baselines. Local information helps RankMCL to beat RankMC. PTR4EI, besides local features, exploits complementary context information and particular Twitter features, such as the presence of hash-tags or URLs in the tweets, this information allows it to improve its ranking performance even further, reaching a P@10 of 96\%. A similar behavior is observed for MAP and NDCGL~\cite{ICWSMDIAZ2012}.
%

\subsection{Lessons Learned and Outlook}
After the detection of the outbreak, authorities investigating the cause and impact of the outbreak are interested in the analysis of micro-blog data related to the event. Millions of health-related tweets are produced every day, which make this task overwhelming for the experts. Yet, our approach, PTR4EI demonstrated a superior ranking performance and was able to provide users with a personalized short list of tweets that met the context of their investigation. PTR4EI exploits features that go beyond the medical condition and location (i.e., user context) and includes complementary context information, extracted using LDA and the social hash-tagging behavior in Twitter.

The main advantage of PTR4EI is that it can discover new relationships from the dynamic data stream based on a limited context in order to help filtering the large amount of data. 

The method presented here requires labeled data to train the model; placing extra effort and burden on experts. We are currently considering crowdsourcing as a complementary means to obtain labeled data on demand, similar to the crowdsourcing approach discussed in Stage~I (Section~\ref{sec:approach3}).
\section{Related Works on Epidemic Intelligence from Social Media}
\label{sec:realted}
In recent years there has been significant research efforts in analyzing tweets to enhance outbreak alerts, which can urge a rapid response from health authorities, thereby helping them to prevent and/or mitigate public health threats. Here we present some illustrative instances.

In the case of Influenza-like Illnesses (ILI), for example,  Culotta~\cite{Culotta:2010:TDI:1964858.1964874} analyzes tweets to determine if influenza-related messages correlates with influenza statistics reported by the Centers for Disease Control and Prevention (CDC) in the United States. The author found a positive correlation with the official statistics. 


Lampos and Cristianini~\cite{LamposTIST2011} address the task of detecting the diffusion of ILI from tweets. Their analysis uses a statistical learning framework based on LASSO and L1-norm regularization in order to select a consistent subset of textual features from a large amount of candidates. They observe that their approach is able to select features with close semantic correlation with the target health related topics and that the regression models have a significant performance improvement.

Beyond ILI, tweets time series and user behavior also have been analyzed to enhance outbreak alerts for other diseases. For example, in~\cite{Chunara01012012} and ~\cite{GomideVMABFT2011} the authors monitor Twitter to understand and characterize Cholera and Dengue outbreaks, respectively.   
%

The aforementioned studies mostly focus on individual countries with a high density of Twitter users, e.g., United Sates, the United Kingdom, or Brazil, and none of them has focused on more than two simultaneously diseases for outbreak detection. Thus, even if they show the advantage of using Twitter for detecting real world outbreak events, they do not consider the temporal dynamics of tweets regarding different diseases in different countries, as we do in this work\footnote{The study in \cite{cikm:kanhabua} goes in this direction, but it does not provide a quantitative correlation between tweets and real word outbreaks.}. 

Note that there are existing EI systems such as the BioCaster Global Health Monitor\footnote{\url{http://biocaster.nii.ac.jp/}}, or HealthMap\footnote{\url{http://www.healthmap.org/en/}}. However, they differ from our proposed system in the level of analyses and data mining models, information sources, and results presentation and visualization. Furthermore, \meco's\ personalization and filtering techniques are key differentiators of our approach.

\section{Conclusion}
\label{sec:conclusions}
Leveraging social media for Epidemic Intelligence systems is a promising but also a challenging endeavor.

For message filtering, the main challenge lies in the ambiguity of term usage
and of terminology evolution in Twitter. We find that semi-supervised classification works well when using labeled data for training the classifier and retraining on feature changes. Our proposed algorithm for detecting novel tweets can identify such feature changes and select a sample of corresponding messages for human assessment.

Interestingly, it is not necessary to let medical experts label the training data;
with crowdsourcing, a similar level of labeling quality could be achieved.
As the main aim of this stage is high recall, and false positives are acceptable,
a supervised classifier trained with a regularly updated, crowd-labeled training
set, is a feasible solution.

With respect to alert generation, we identified four different classes of time series data, based on the two characteristics: oscillation and magnitude. For low volume cases, messages can be directly treated as alerts because the cognitive load for later assessment is small. Low oscillation and high magnitude cases feature pronounced message peaks for outbreaks; this type of event is easily detected by biosurveillance algorithms. The challenging type is the class of high oscillation and high magnitude time series. 
%

We can conclude that while many time series are amenable to reliable alert generation, for particular cases (e.g., high oscillation and high magnitude time series) more research is needed to devise algorithms which are more robust under noise and incomplete data.

For detected events, public health experts face the overwhelming task of analyzing the large number of tweets associated to the alerts. In order to reduce this information overload and support the task of threat assessment, we leveraged complementary context information discovered within the tweets --~i.e., extracted from the social hash-tagging and latent topics. We were able to achieve an effective ranking mechanism for messages associated with alerts.

To summarize, available techniques are sufficiently mature to build useful monitoring and early warning systems based on social media streams. Collective intelligence can be employed not only as a valuable information source, but also for tasks such as training data creation. Further work is needed to devise alert generation algorithms with a better recall-precision trade-off. However, the current load of experts in assessing these alerts can be reduced significantly by employing personalized ranking techniques. 

We are confident that this study brings Epidemic Intelligence based on social media a step forward and we hope it provides insightful lessons for similar ventures.
\begin{acks}
This work was partially funded by the European Commission Seventh Framework
Program (FP7 / 2007-2013) under grant agreement No.247829 for the Medical Ecosystem Project (M-Eco). We thank the epidemiologists at the Nieders\"achsisches Landesgesundheitsamt and the Robert-Koch Institute for providing domain expert recommendations and advice for this work.
\end{acks}
\clearpage
\bibliographystyle{ACM-Reference-Format-Journals}
\bibliography{all,ernesto}

\received{XXXXX}{XXXXX}{XXXXX}

%
%
%
%
%
%

\end{document}